\documentclass[preprint,showkeys,showpacs,preprintnumbers,amsmath,amssymb,tightenlines,letter]{revtex4}

\usepackage{amsfonts}
\usepackage{graphicx}
\usepackage{natbib}
\usepackage{dcolumn}
\usepackage{bm}
\usepackage{rotating}

\begin{document}

\preprint{}

\title{On approximating the distributions\\ of goodness-of-fit test statistics\\
based on the empirical distribution function:\\ The case of
unknown parameters}

\author{Marco Capasso}
\email{marco.capasso@sssup.it}

\author{Lucia Alessi}
\email{lucia.alessi@sssup.it}

\author{Matteo Barigozzi}
\email{matteo.barigozzi@sssup.it}

\author{Giorgio Fagiolo}
 \email{giorgio.fagiolo@sssup.it}
\affiliation{Sant'Anna School of Advanced Studies, Laboratory of
Economics and Management, Pisa, Italy.}

\bigskip

\date{April 2008}

\bigskip

\begin{abstract}
\noindent This paper discusses some problems possibly arising when
approximating via Monte-Carlo simulations the distributions of
goodness-of-fit test statistics based on the empirical distribution
function. We argue that failing to re-estimate unknown parameters on
each simulated Monte-Carlo sample -- and thus avoiding to employ
this information to build the test statistic -- may lead to wrong,
overly-conservative testing. Furthermore, we present a simple
example suggesting that the impact of this possible mistake may turn
out to be dramatic and does not vanish as the sample size increases.
\end{abstract}



\maketitle

\section{Introduction} \label{Section:Introduction}

This paper discusses some problems possibly arising when
approximating -- via Monte-Carlo simulations -- the distributions
and critical values of the most commonly employed goodness-of-fit
(GoF) tests based on empirical distribution function (EDF)
statistics \cite{DagoSteph1986,Thode2002}.

This situation arises very frequently -- in many areas of
statistical physics or econophysics -- when the researcher aims at
fitting some experimental or empirical (univariate) sample with a
parametric (univariate) probability distribution whose parameters
are unknown. In such cases, the goodness of fit may be ex-post
evaluated by employing standard statistical tests based on the EDF.
If, as typically happens, critical-value tables are not available,
one has to resort to Monte-Carlo methods to derive the approximated
distribution of the test statistics under analysis.

We show that, when testing with unknown parameters, critical values
(and consequently testing outcomes) may be dramatically sensible to
the details of the Monte-Carlo procedure actually employed to
approximate them. More specifically, we argue that the researcher
may sometimes build inaccurate critical-value tables because he/she
fails to perform a crucial step in his/her Monte-Carlo simulation
exercises, namely maximum-likelihood (ML) re-estimation of unknown
parameters on each simulated sample. In our opinion, this is a
lesson worth learning because critical-value tables are only
available for particular distributions (e.g., normal, exponential,
etc.). In all other cases, our study indicates that failing to
correctly specify the Monte-Carlo approximation procedure may lead
to overly-conservative hypothesis tests.

The rest of this paper is organized as follows. Section 2 formalizes
the general GoF test under study and discusses the main problems
associated to the approximation of EDF-based GoF test-statistic
distributions from a theoretical perspective. Section 3 presents an
application to the case of normality with unknown parameters.
Finally, Section 4 concludes with a few remarks.

\section{Approximating EDF-based GoF test-statistic
distributions} \label{Sec:TheProblem}

In many applied contexts, the researcher faces the problem of
assessing whether an empirical univariate sample
$\underline{x}_N=(x_1,\dots,x_N)$ comes from a (continuous)
distribution $F(x;\theta)$, where $\theta$ is a vector of
\textit{unknown} parameters. EDF-based GoF tests
\cite{DagoSteph1986,Thode2002} employ statistics that are non-
decreasing functions of some distance between the theoretical
distribution under the null hypothesis $H_0: \underline{x}_N \sim
F(x;\theta)$ and the empirical distribution function constructed
from $\underline{x}_N$, provided that some estimate of the unknown
parameters is given.

In what follows, we will begin by focusing on the simplest case
where $F(x;\theta)$ has only location and scale unknown parameters
(we will discuss below what happens if this is not the case).
Furthermore, we will limit the analysis to four out of the most used
EDF test statistics, namely Kolmogorov-Smirnov
\cite{Massey1951,Owen1962}, Kuiper \cite{Kuiper1962}, Cram\'{e}r -
Von Mises \cite{PearsonStephens1962} and Quadratic Anderson- Darling
\cite{AndersonDarling1954}, with small-sample modifications usually
considered in the literature \footnote{For more formal definitions,
see \cite{DagoSteph1986}, Chapter 4, Table 4.2. Small-sample
modifications have been applied to benchmark our results to those
presented in the literature. However, our main findings remain
qualitatively unaltered if one studies test-statistic distributions
without small-sample modifications.}.

It is well-known that if one replaces $\theta$ with its maximum
likelihood (ML) empirical-sample estimate
$\hat{\theta}(\underline{x}_N)$, the distributions of the EDF test
statistic under study can be shown to be independent on the unknown
true parameter values \cite{DavidJohnson1948}. However,
test-statistic distributions are hard to derive analytically. They
must be therefore simulated via Monte-Carlo and critical values must
be accordingly computed. To do so, let us consider a first possible
procedure:

\begin{center} \bf{Procedure A}
\end{center}

\begin{description}

\item[\textbf{Step A1}] \it{Generate, by means of standard simulation
techniques \cite{Devroye1986}, a sufficiently large number (say,
$M>>0$) of independently-drawn N-sized samples
$\underline{z}_N^j=(z_1^j,\dots,z_N^j)$, $j=1,\dots,M$, where each
$z_i^j$ is an i.i.d. observation from a
$F(x;\hat{\theta}(\underline{x}_N))$, i.e. from the distribution
under $H_0$ where unknown parameters are replaced by their
empirical-sample estimates;} \bigskip

\item[\textbf{Step A2}] \it{For each N-sized sample $\underline{z}_N^j$,
compute an observation of the EDF test statistic under study by
comparing the EDF constructed from $\underline{z}_N^j$ with the
theoretical distribution\\
$F(\underline{z}_N^j,\hat{\theta}(\underline{x}_N))$, i.e. when $F$
is computed at the empirical sample observations and} \emph{unknown
parameters are always replaced with estimates
$\hat{\theta}(\underline{x}_N)$ obtained once and for all from the
empirical sample}; \bigskip

\item[\textbf{Step A3}] \it{Once Step A2 has been carried out for all $M$ samples, compute the
empirical distribution function $T$ of the test statistic;} \bigskip

\item[\textbf{Step A4}] \it{Compute (upper-tailed) critical values, for any given significance
level $\alpha$, by employing the empirical distribution function $T$
of the EDF test statistic as obtained in Step A3. }

\end{description}

 At a first scrutiny, the above procedure seems to be
correct. Indeed, the procedure tells us to approximate the
distribution of the test statistic under study by repeatedly compute
it on a sufficiently large number of i.i.d. samples, all distributed
as if they came from the null distribution $F(\cdot,\theta)$, when
the unknown parameters are replaced with their empirical sample
estimate $\hat{\theta}(\underline{x}_N)$.

Despite its appeal, however, Procedure A can be shown to be wrong,
in the sense that it generates a completely wrong approximation to
the ``true'' distribution of the test statistic under the null
hypothesis.

The reason why Procedure A is not correct lies in Step A2. More
precisely, when we compare the EDF constructed from
$\underline{z}_N^j$ with the theoretical distribution
$F(\underline{z}_N^j,\hat{\theta}(\underline{x}_N))$, we are
assuming that our estimate for $\theta$ does not depend on the
actual sample $\underline{z}_N^j$ under analysis. This is the same
as presuming that the hypothesis test is performed for \textit{known
parameters}. On the contrary, sticking to the null hypothesis
implies that the theoretical distribution which should be compared
to the EDF of $\underline{z}_N^j$ must have parameter estimates that
depend on the actual Monte-Carlo sample $\underline{z}_N^j$. In
other words, scale and location parameters $\theta$ must be
re-estimated (via, e.g., ML) \textit{each time we draw the
Monte-Carlo sample}. Let $\hat{\theta}(\underline{z}_N^j)$ be such
estimate for sample $j$. This means that the theoretical
distribution to be used to compute the test statistic would be
$F(\underline{z}_N^j,\hat{\theta}(\underline{z}_N^j))$ and not
$F(\underline{z}_N^j,\hat{\theta}(\underline{x}_N))$. The correct
procedure therefore reads:

\newpage
\begin{center} \bf{Procedure B}
\end{center}

\begin{description}

\item[\textbf{Step B1}] \it{Same as A1}; \bigskip

\item[\textbf{Step B2}] \it{For each N-sized sample $\underline{z}_N^j$,
compute an observation of the EDF test statistic under study by
comparing the EDF constructed from $\underline{z}_N^j$ with the
theoretical distribution
$F(\underline{z}_N^j,\hat{\theta}(\underline{z}_N^j))$, i.e. when
$F$ is computed at the empirical sample observations and}
\emph{unknown parameters are replaced with estimates
$\hat{\theta}(\underline{z}_N^j)$ obtained from the j-th Monte-Carlo
sample}; \smallskip

\item[\textbf{Step B3}] \textit{Same as A3}; \bigskip

\item[\textbf{Step B4}] \textit{Same as A4}.

\end{description}

How dramatic is the error we make in applying Procedure A instead of
Procedure B? Do we get a more conservative or less
conservative\footnote{We loosely define here a test statistic to be
``more conservative'' if it allows to accept the null hypothesis
with a higher likelihood, given any significance levels.} test by
using the wrong procedure? In other words, can we detect significant
shifts in the Monte-Carlo approximation to the distribution of the
test statistics under study when we compare Procedures A and B? In
the next section, we will answer these questions by providing a
simple example.

\section{Application: Testing for Normality with Unknown
Parameters} \label{Sec:Application}

Let us consider the null hypothesis that the empirical sample comes
from a normal distribution $N(\mu,\sigma)$ with unknown mean ($\mu$)
and standard deviation ($\sigma$). In such a case, parameters may be
replaced by their ML estimates
$(m(\underline{x}_N),s(\underline{x}_N))$, i.e. sample mean and
standard deviation. In this case critical values for the four test
statistics under study are already available. Our goal, for the sake
of exposition, is therefore to compare Monte-Carlo approximations to
the distributions of the four test statistics obtained under
Procedures A and B.

We thus have two setups. In the first one (Procedure A), one does
not re-estimate the parameters and always employs
$(m(\underline{x}_N),s(\underline{x}_N))$ to build the theoretical
distribution. In the second one (Procedure B), one re-estimates via
ML mean and standard deviation on each simulated sample by computing
$(m(\underline{z}_N^j),s(\underline{z}_N^j))$ and then uses them to
approximate the theoretical distribution of the test statistic.

Our simulation strategy is very simple. Since the argument put forth
above does not depend on the observed sample's mean and standard
deviation, we can suppose that
$(m(\underline{x}_N),s(\underline{x}_N))=(0,1)$ without loss of
generality\footnote{Alternatively, one can standardize the observed
sample and generate Monte-Carlo sample replications from a N(0,1)
without loss of generality.}. For each of the four test statistics
considered, we run Monte-Carlo simulations \footnote{All simulations
are performed using MATLAB$^{\circledR}$, version 7.4.0.287
(R2007a).} to proxy its distribution under the two setups above. In
both setups, we end up with an approximation to the distribution of
the four tests, from which one can compute critical values
associated to any significance level (or p-value).

To begin with, Table \ref{bigtable} shows critical values for all 4
tests at $\alpha=0.05$ significance level, and for different
combinations of $N$ (sample size) and $M$ (Monte-Carlo
replications). It is easy to see that if we employ Procedure B, we
obtain the same critical values published in the relevant literature
for the case of normality with \textit{unknown} parameters (compare,
e.g., our table \ref{bigtable} with table 1A-1.3 at page 732 in
Stephens, 1974\nocite{Stephens1974}). On the contrary, if we employ
procedure A, critical values dramatically increase. The effect is of
course more evident in the case of so-called ``quadratic
statistics'' (Cram\'{e}r-Von Mises and Quadratic Anderson-Darling),
but is equally relevant also in the case of ``supremum statistics''
(Kolmogorov-Smirnov and Kuiper). What is more, Procedure A allows us
to obtain critical-value figures which are very similar to those
found in the literature for the case of normality with
\textit{completely specified, known, parameters}.

\begin{figure*}[p]
        \begin{minipage}[h]{7cm}
        \begin{scriptsize}
        \centering {\includegraphics[width=7cm]{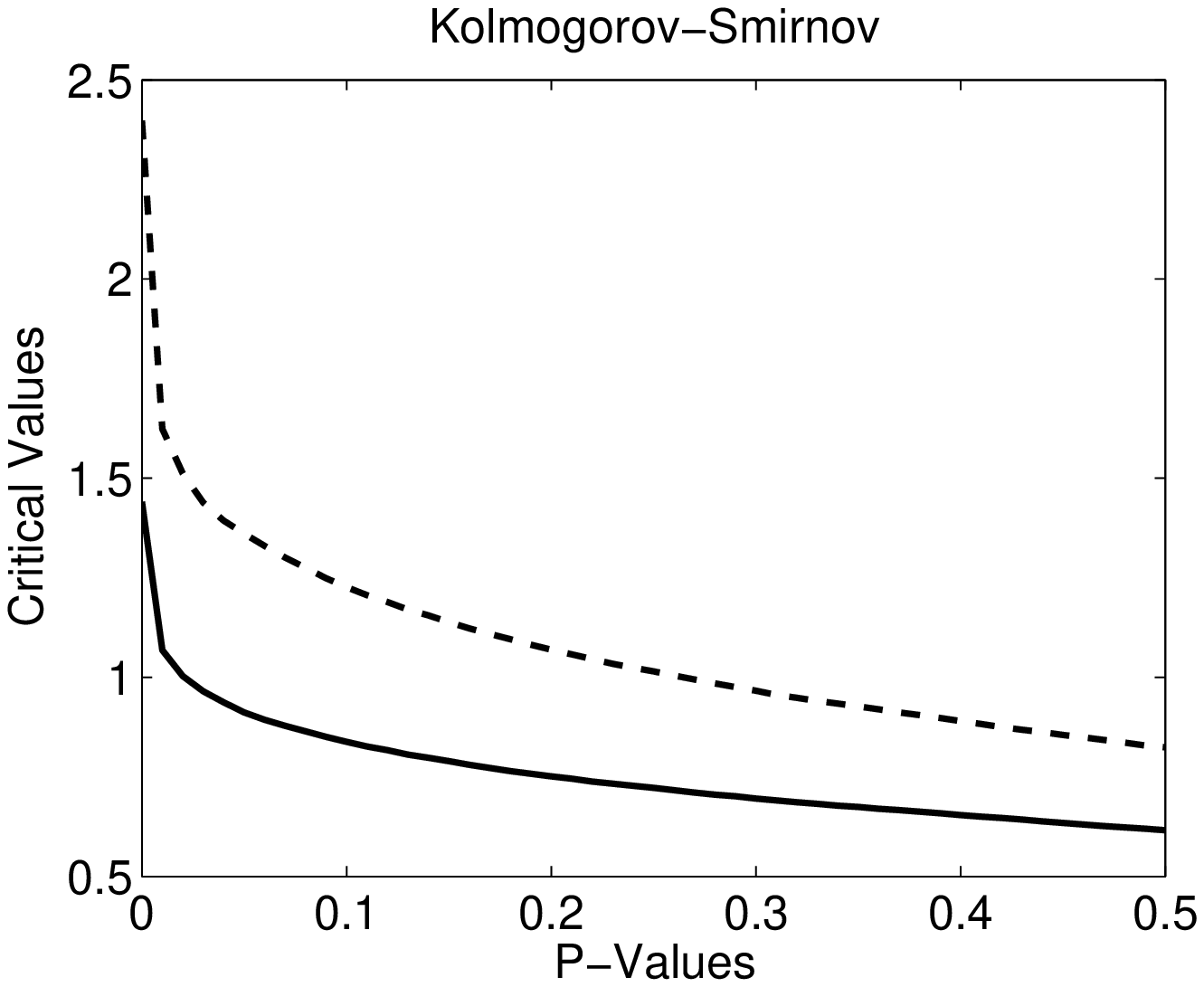}}
        \end{scriptsize}
        \end{minipage}\hfill
        \begin{minipage}[h]{7cm}
        \begin{scriptsize}
        \centering {\includegraphics[width=7cm]{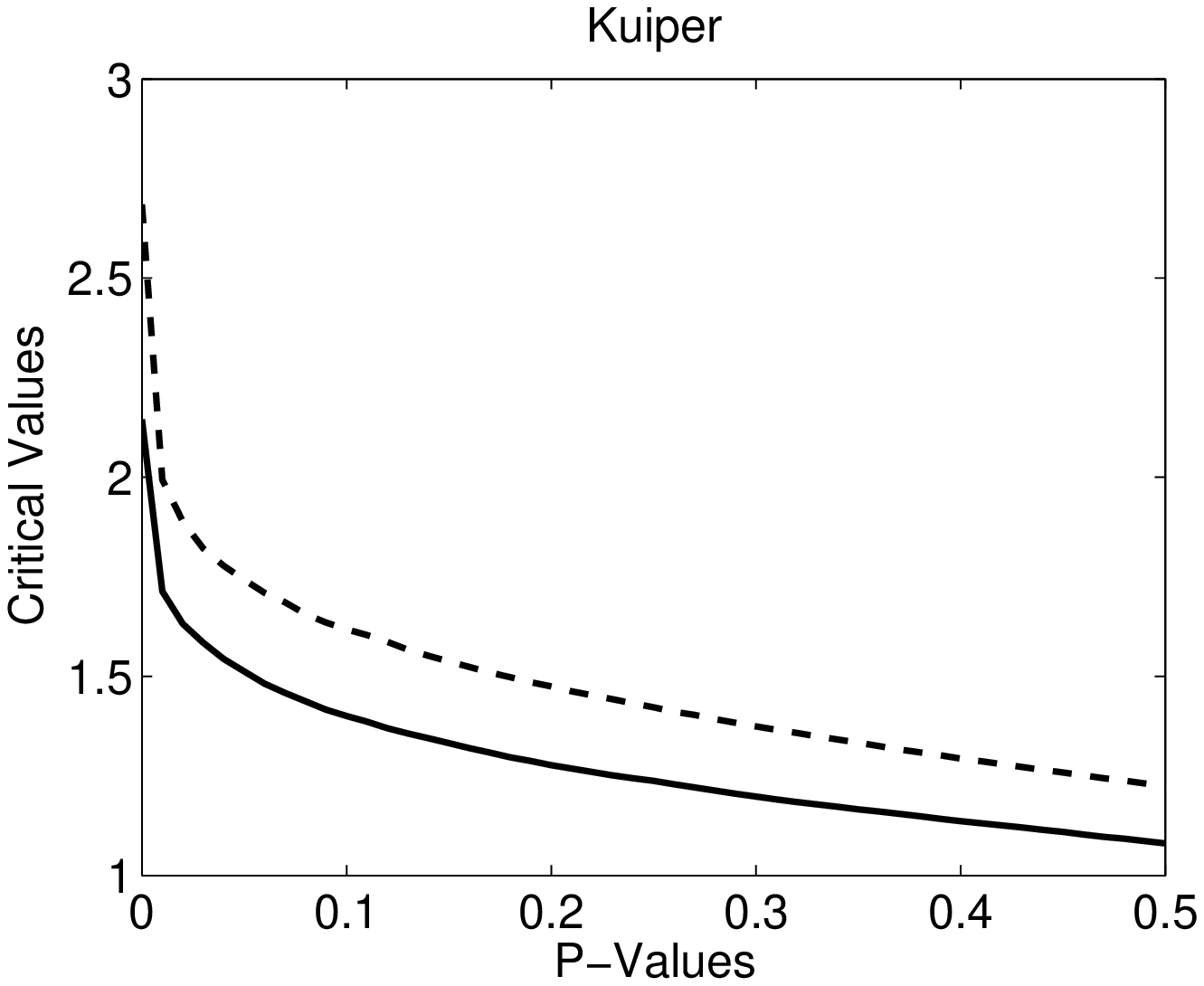}}
        \end{scriptsize}
        \end{minipage}
        \begin{minipage}[h]{7cm}
        \begin{scriptsize}
        \centering {\includegraphics[width=7cm]{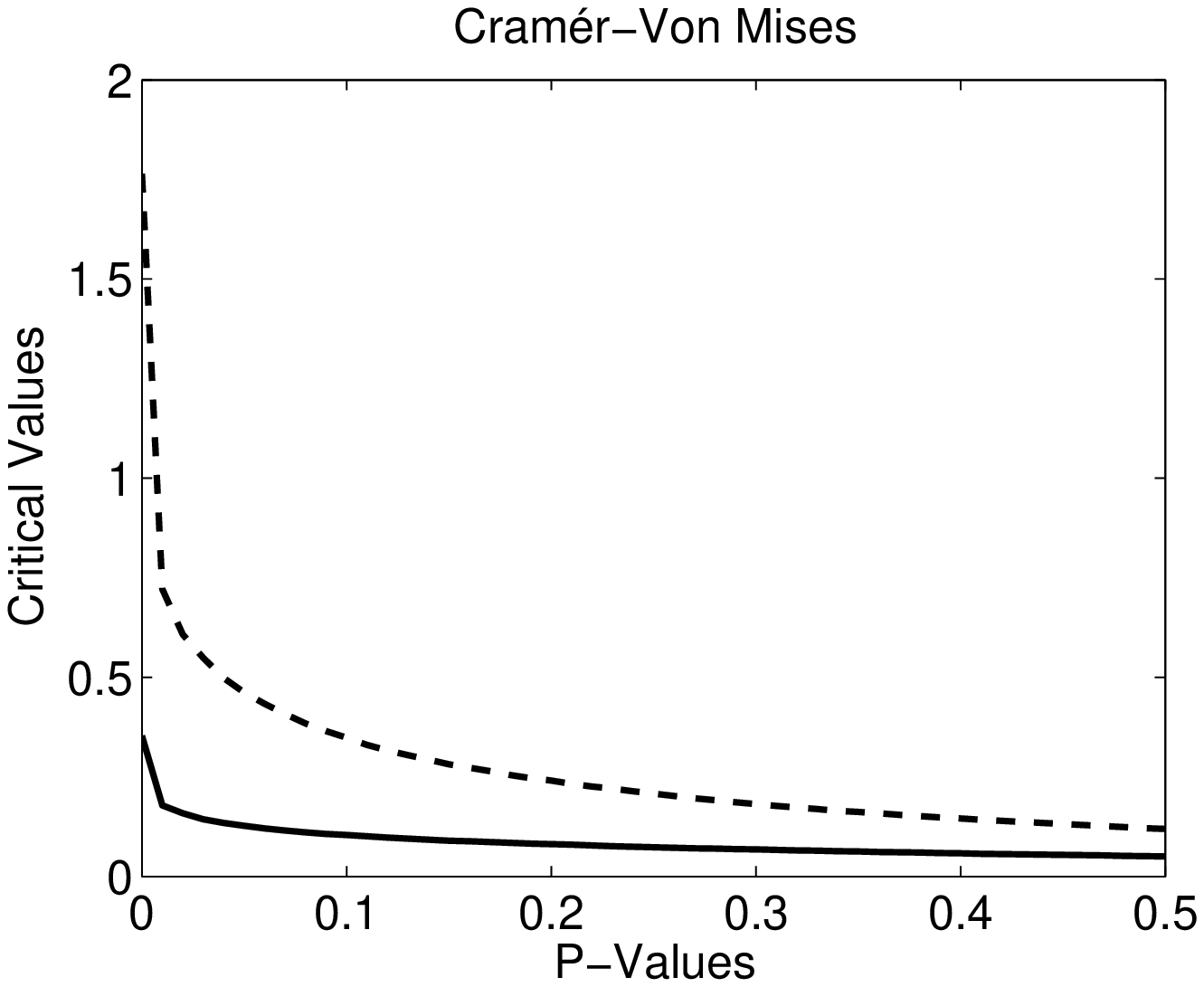}}
        \end{scriptsize}
        \end{minipage}\hfill
        \begin{minipage}[h]{7cm}
        \begin{scriptsize}
        \centering {\includegraphics[width=7cm]{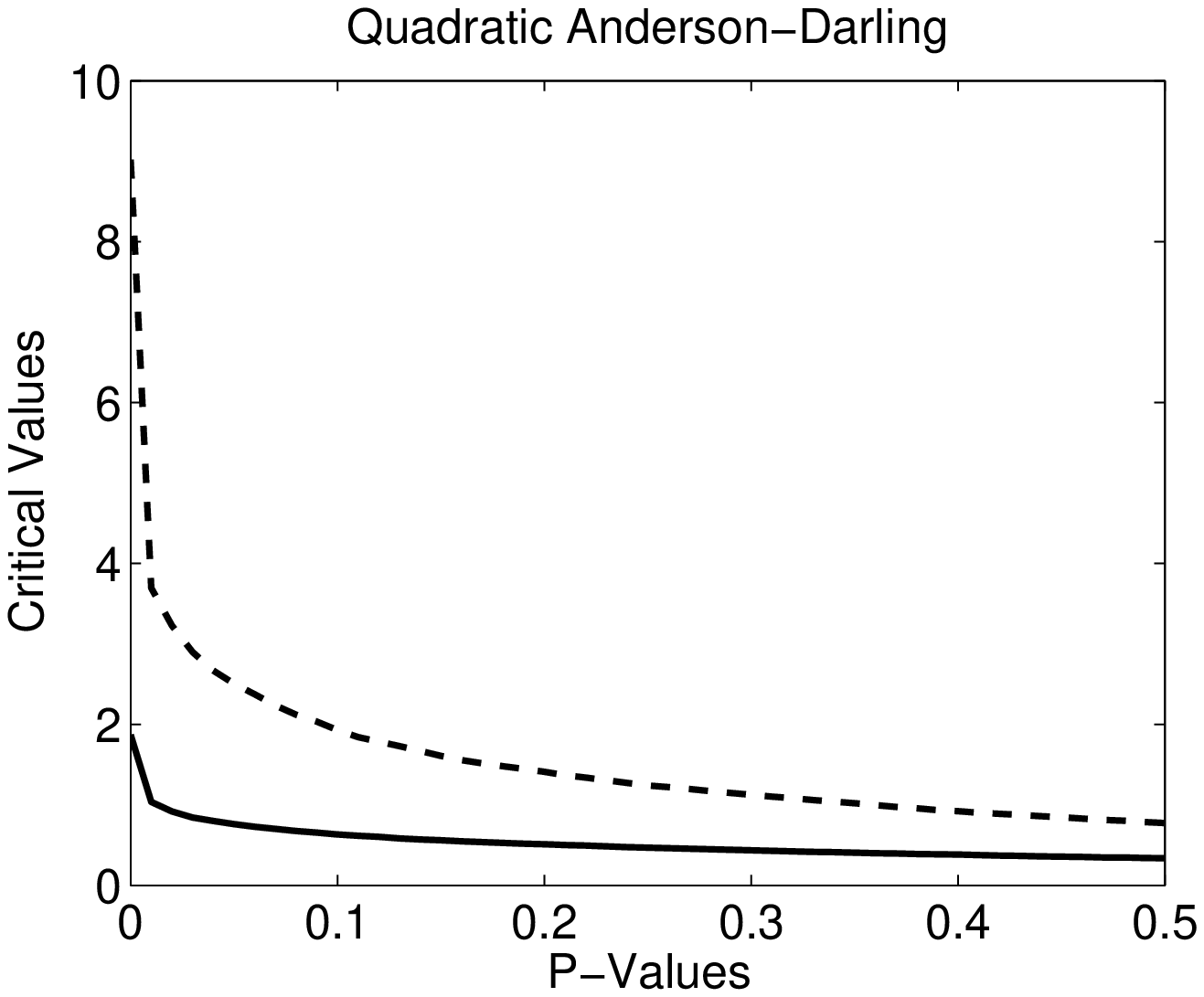}}
        \end{scriptsize}
        \end{minipage}
        \caption{Critical values versus P-values for the four test statistics under
        study. Empirical sample size: $N=5000$. Number of Monte-Carlo replications: $M=10000$.
        Solid line: Procedure B (parameters are re-estimated each time on Monte-Carlo sample).
        Dashed line: Procedure A (always using empirical-sample estimates).}
        \label{pvalues}
        \end{figure*}

\begin{figure*}[p]
        \begin{minipage}[h]{7cm}
        \begin{scriptsize}
        \centering {\includegraphics[width=7cm]{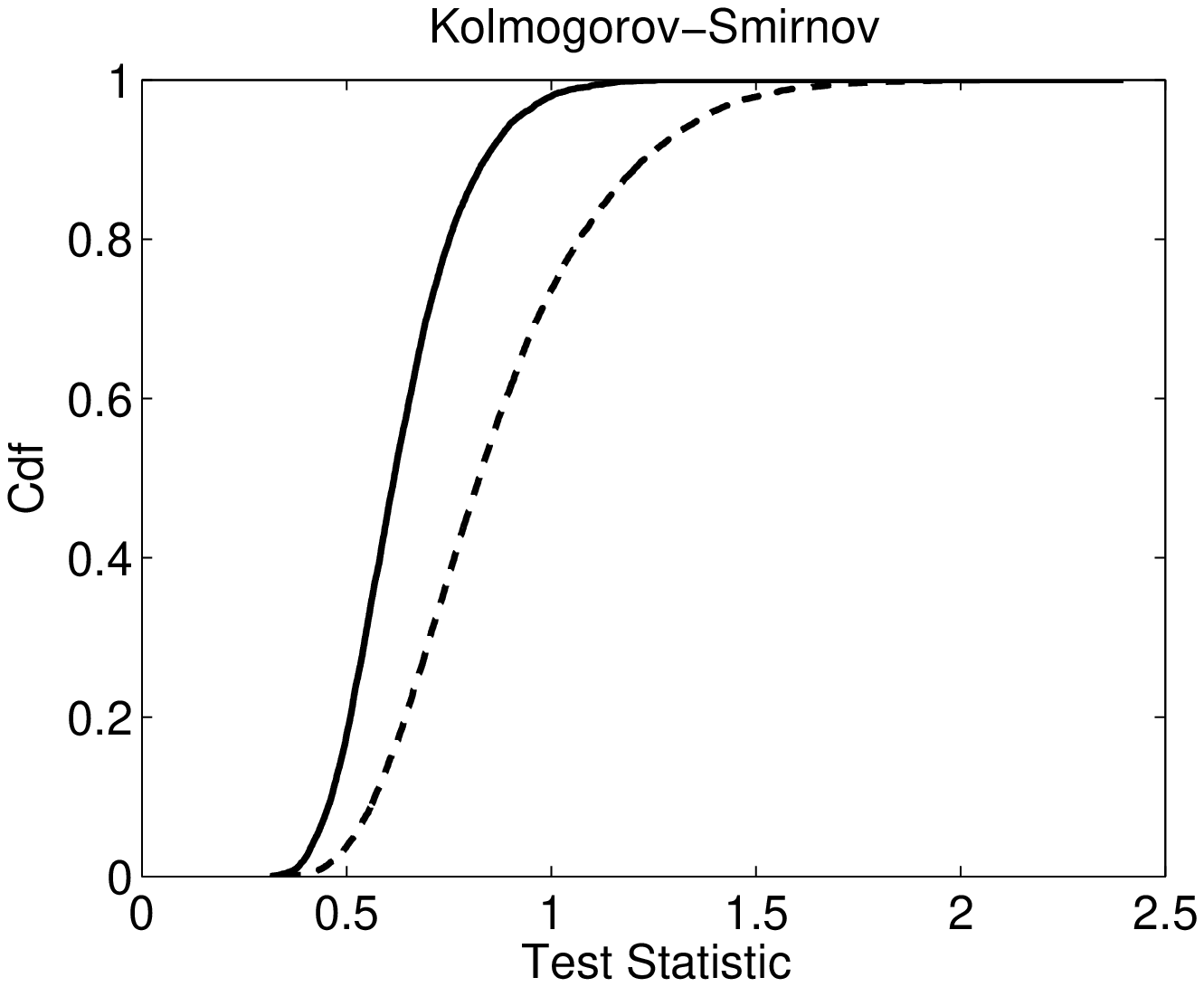}}
        \end{scriptsize}
        \end{minipage}\hfill
        \begin{minipage}[h]{7cm}
        \begin{scriptsize}
        \centering {\includegraphics[width=7cm]{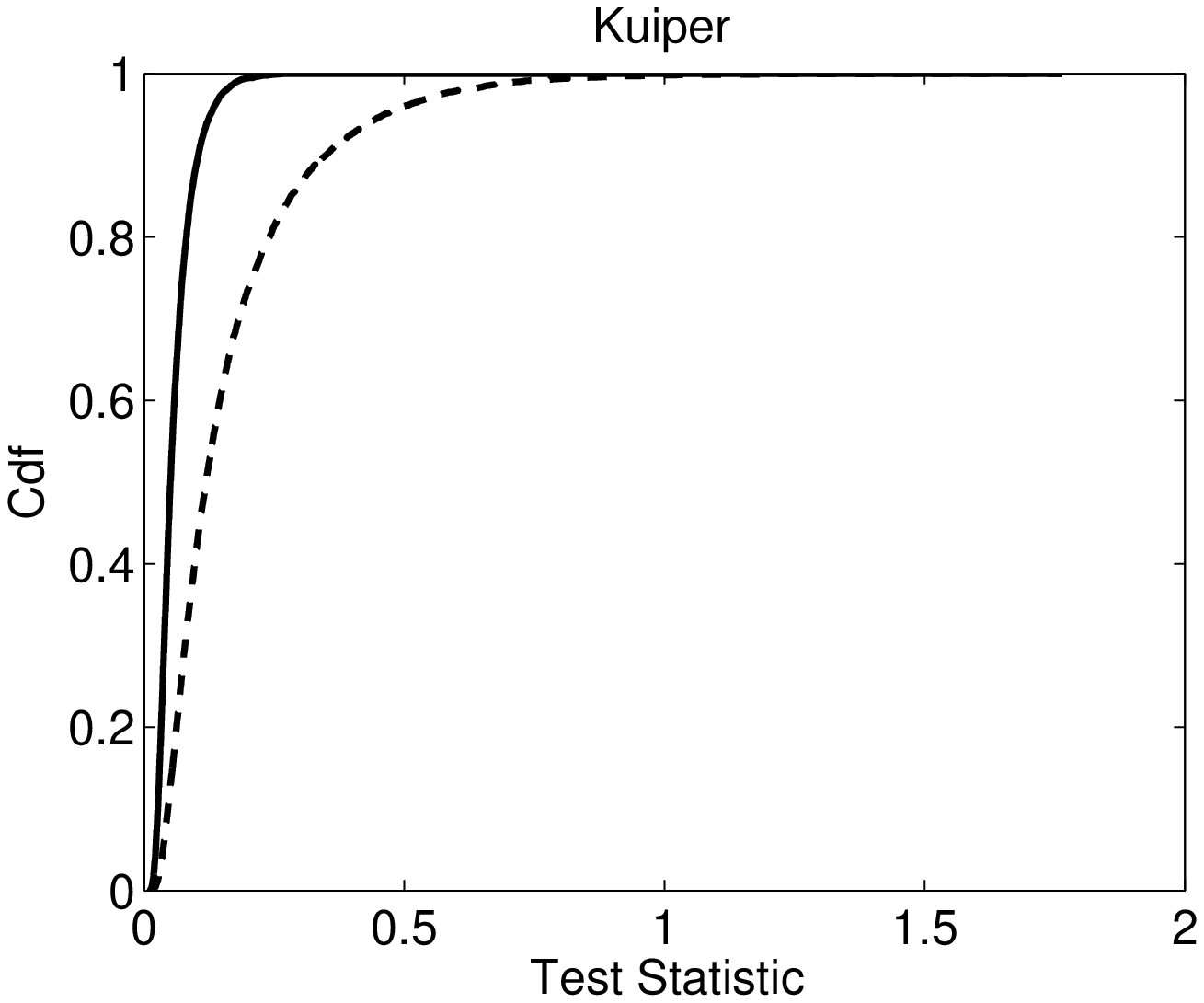}}
        \end{scriptsize}
        \end{minipage}
        \begin{minipage}[h]{7cm}
        \begin{scriptsize}
        \centering {\includegraphics[width=7cm]{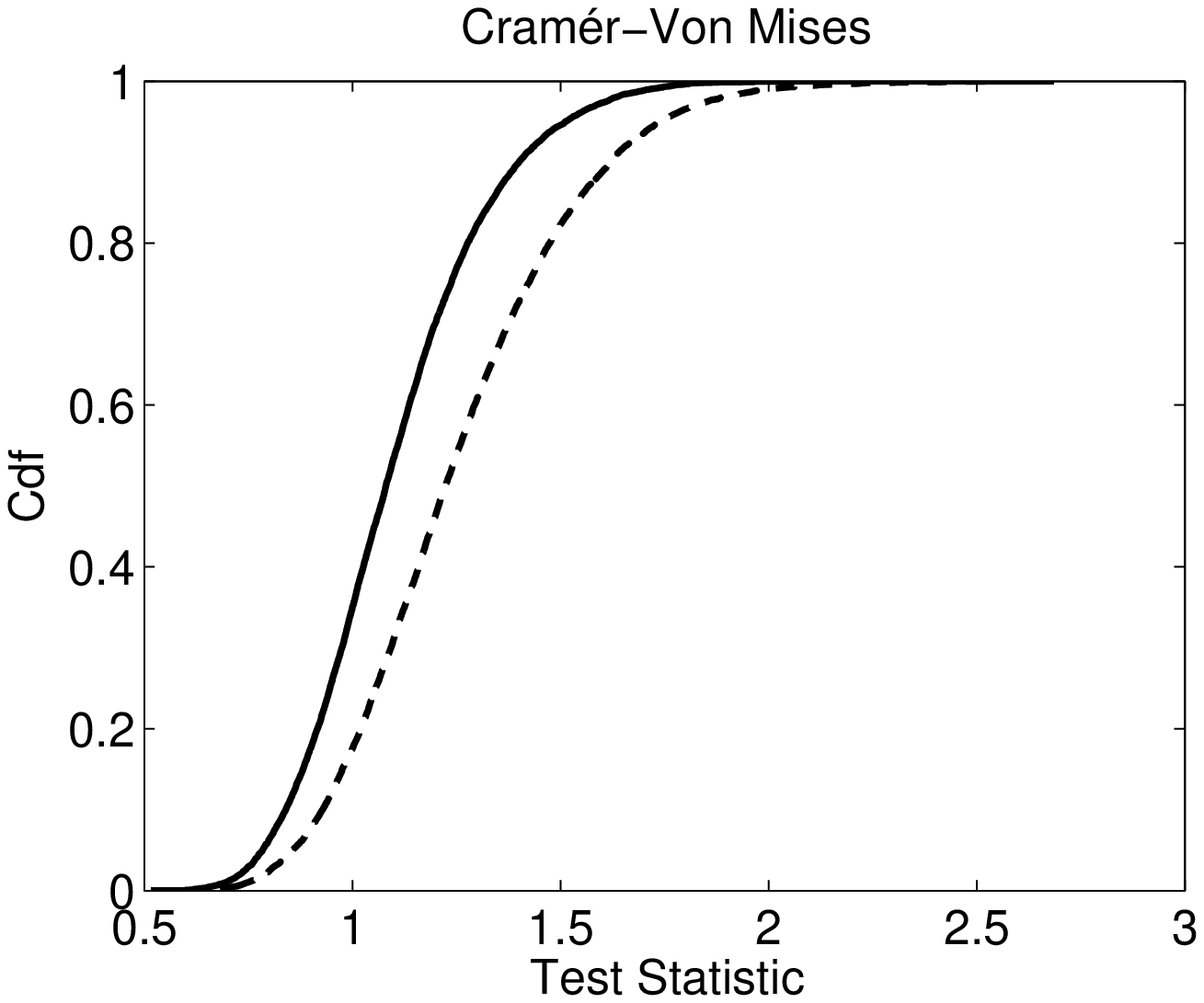}}
        \end{scriptsize}
        \end{minipage}\hfill
        \begin{minipage}[h]{7cm}
        \begin{scriptsize}
        \centering {\includegraphics[width=7cm]{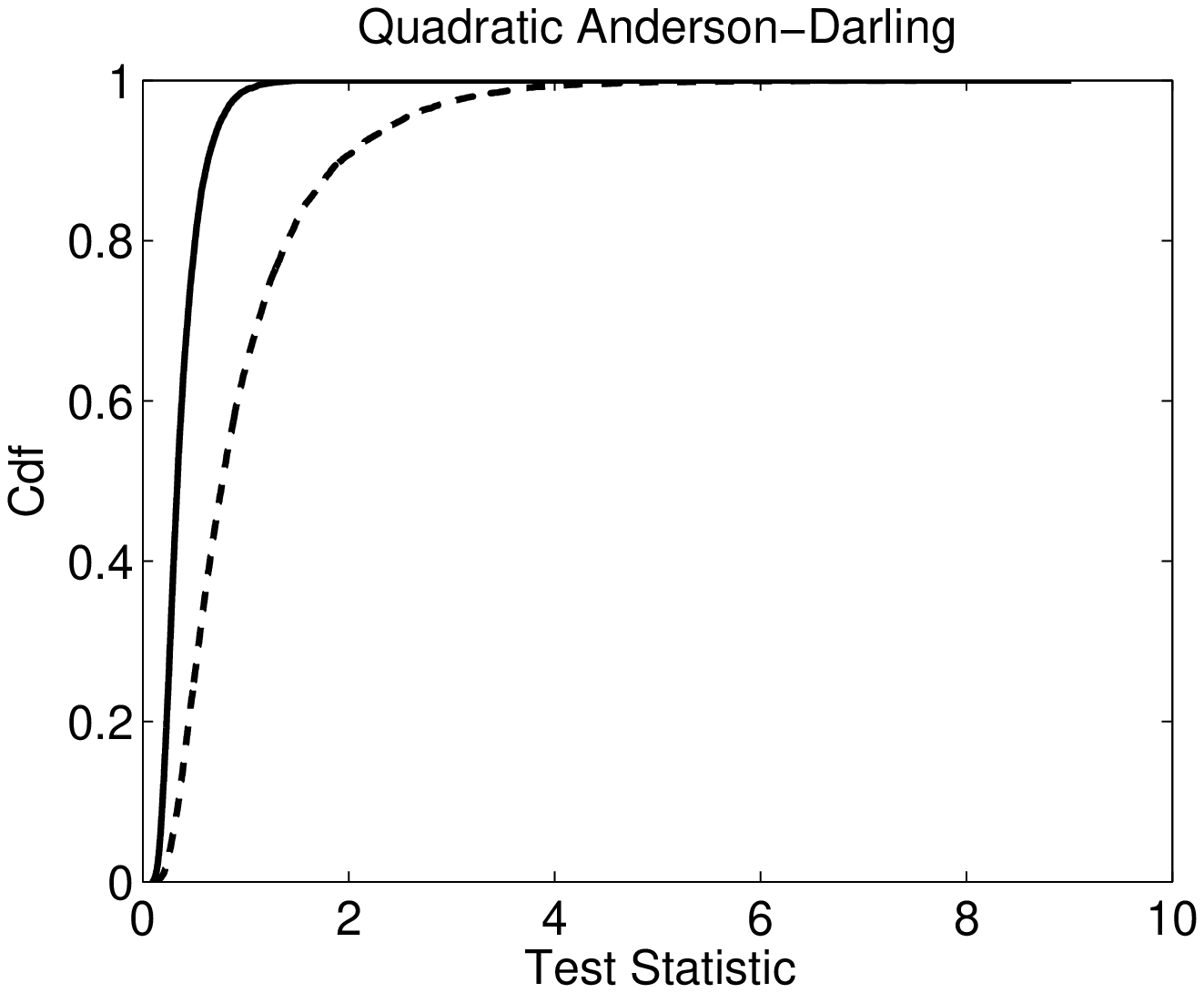}}
        \end{scriptsize}
        \end{minipage}
        \caption{Estimates of cumulative distribution function (Cdf) for the four test statistics under
        study. Empirical sample size: $N=5000$. Number of Monte-Carlo replications: $M=10000$.
        Solid line: Procedure B (parameters are re-estimated each time on Monte-Carlo sample).
        Dashed line: Procedure A (always using empirical-sample estimates).} \label{edistrf}
        \end{figure*}

Table \ref{bigtable} also indicates that if we wrongly employ
Procedure A, we end up with test statistics that are dramatically
more conservative (at $\alpha=0.05$) than if we correctly employ
Procedure B. This is true irrespective of the significance level. As
Figure \ref{pvalues} shows, the A vs. B gap between critical values
remains relevant for all (reasonable) p-value levels. In other
words, the wrong choice of employing Procedure A induces a rightward
shift of (and reshapes) the entire test-statistic distribution. To
see this, in Figure \ref{edistrf} we plot the estimated cumulative
distribution of all 4 test statistics under the two setups. Choosing
Procedure A makes all tests much more conservative.

\begin{figure*}[p]
        \begin{minipage}[h]{7cm}
        \begin{scriptsize}
        \centering {\includegraphics[width=7cm]{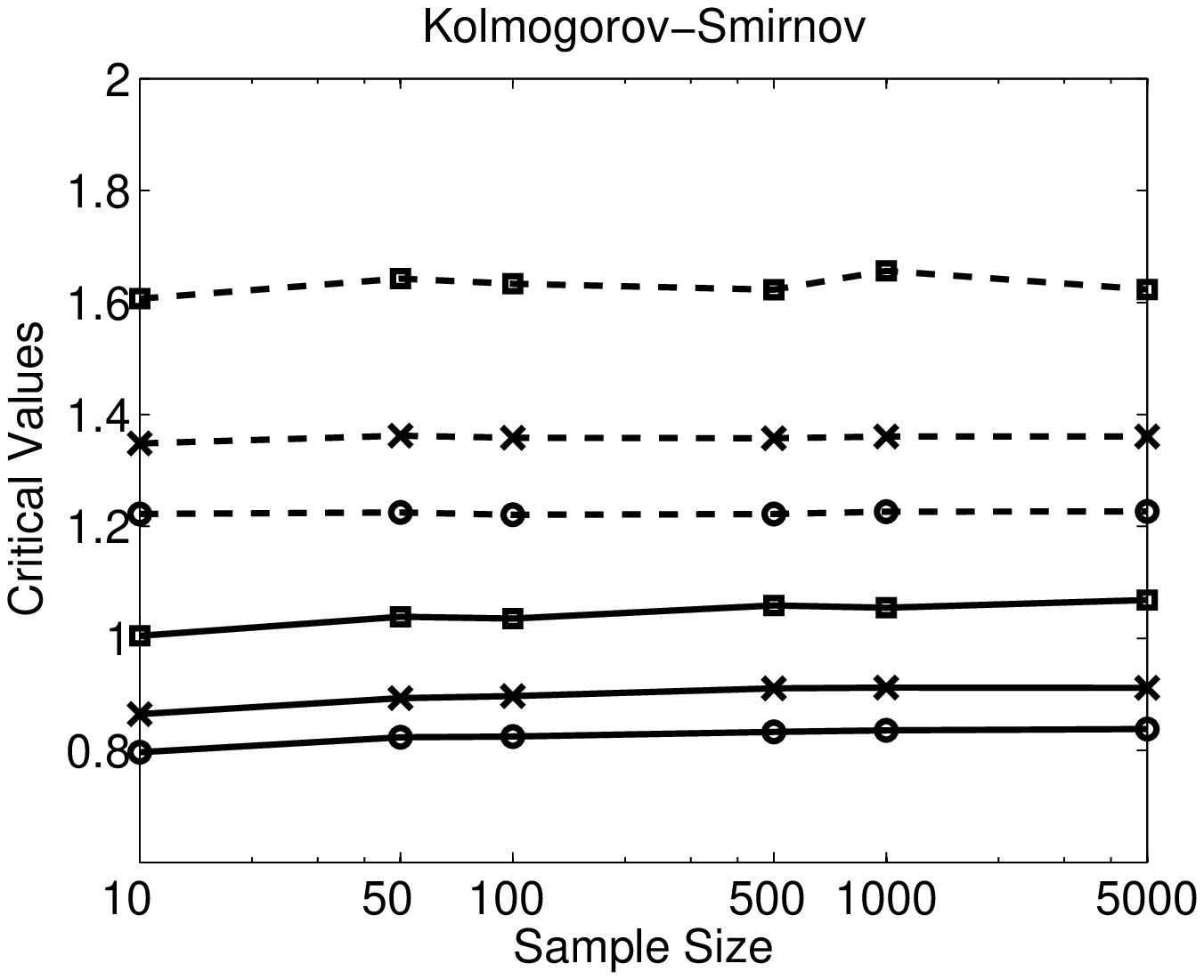}}
        \end{scriptsize}
        \end{minipage}\hfill
        \begin{minipage}[h]{7cm}
        \begin{scriptsize}
        \centering {\includegraphics[width=7cm]{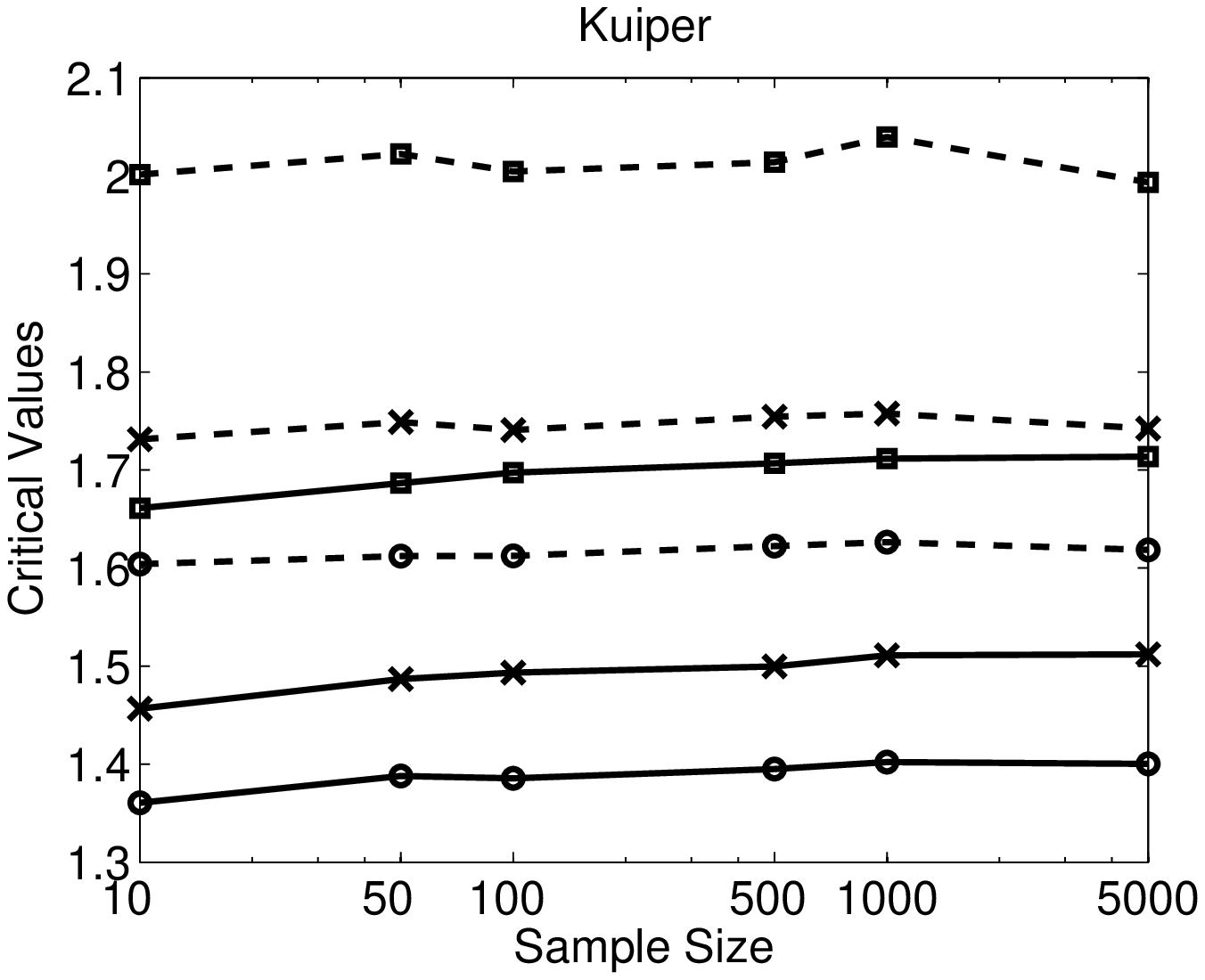}}
        \end{scriptsize}
        \end{minipage}
        \begin{minipage}[h]{7cm}
        \begin{scriptsize}
        \centering {\includegraphics[width=7cm]{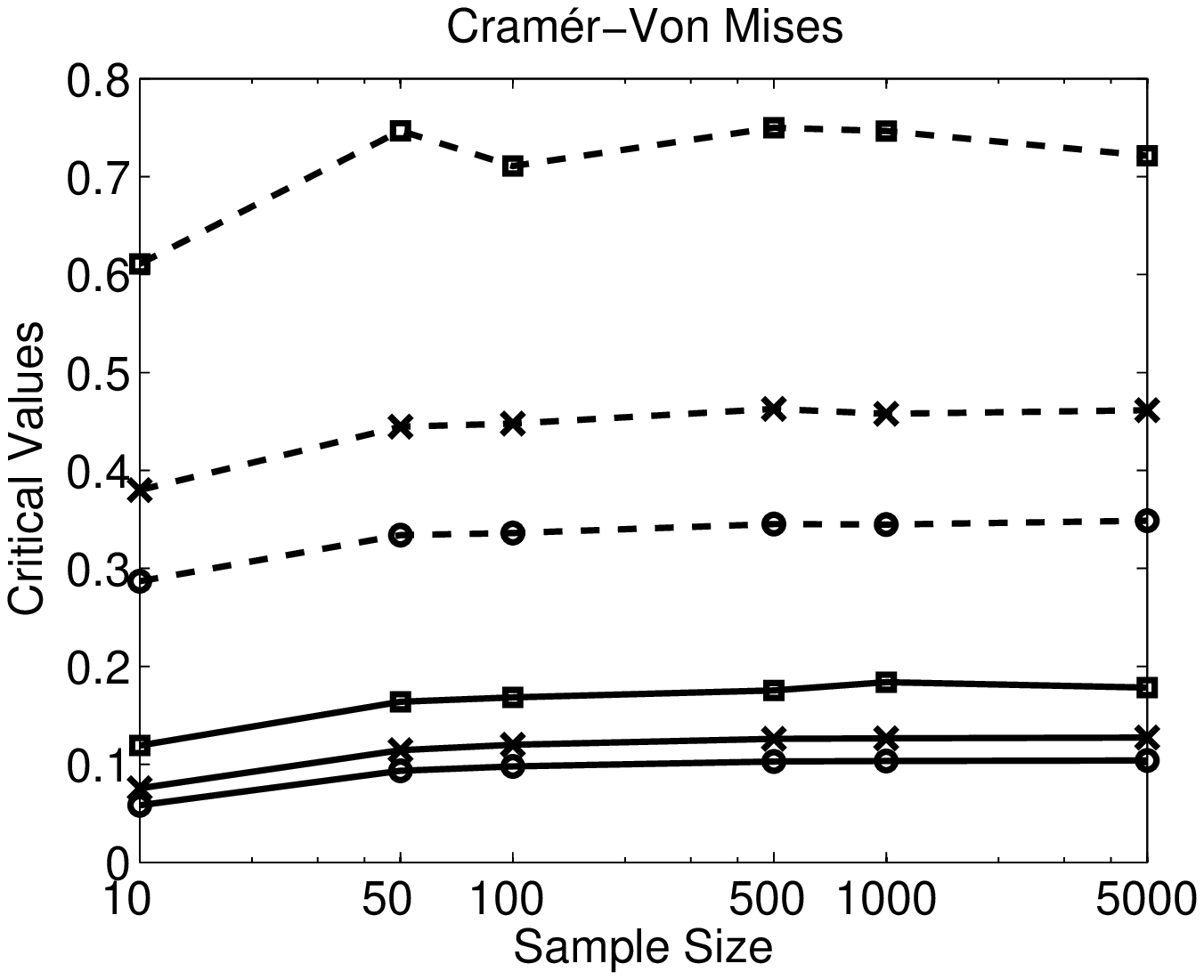}}
        \end{scriptsize}
        \end{minipage}\hfill
        \begin{minipage}[h]{7cm}
        \begin{scriptsize}
        \centering {\includegraphics[width=7cm]{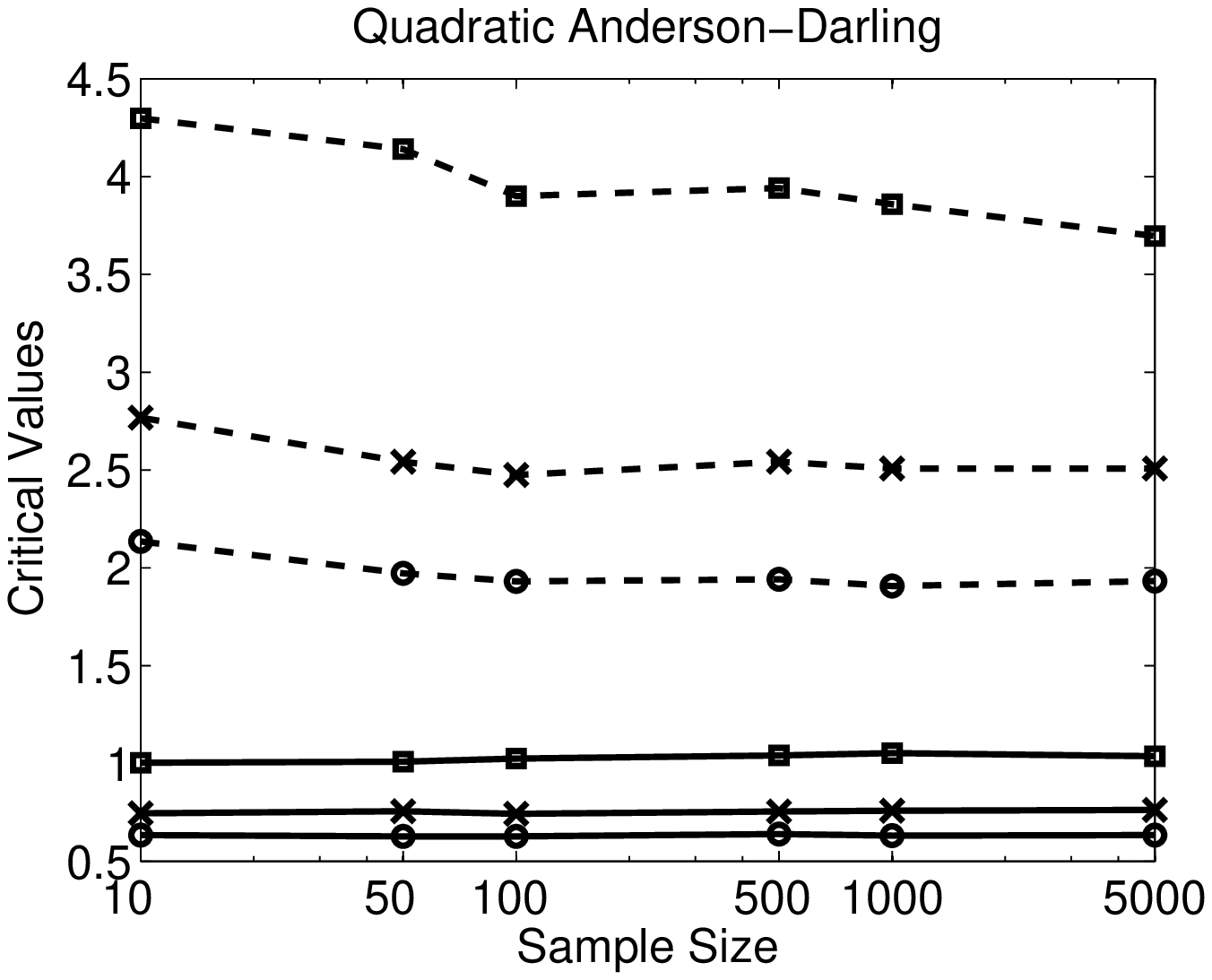}}
        \end{scriptsize}
        \end{minipage}
        \caption{Critical values versus empirical sample size $N$ (in log scale) for the four test statistics under
        study. Number of Monte-Carlo replications: $M=10000$. Solid line: Procedure B (parameters are re-estimated each time on Monte-Carlo sample).
        Dashed line: Procedure A (always using empirical-sample estimates). Symbols stand for significance levels: $\circ$ = 0.10, $\times$ = 0.05, $\square$ = 0.01.}
    \label{changingsample}
    \end{figure*}

Finally, it is worth noting that the above results do not depend on
the empirical sample size. In fact, one might argue that the
mismatch between the two procedures may be relevant only for small
$N$'s but should vanish as $N$ gets large. This is not true: the gap
remains there as $N$ increases within an empirically-reasonable
range and for any sufficiently large number of Monte-Carlo
replications ($M$) -- see Figure \ref{changingsample} for the case
$M=10000$.

\section{Concluding Remarks} \label{Sec:Conclusions}

In this paper, we have argued that failing to re-estimate unknown
parameters on each simulated Monte-Carlo sample (and not employing
this information to compute the theoretical distribution to be
compared with the sample EDF) may lead to wrong, overly-conservative
approximations to the distributions of GoF test statistics based on
the EDF. Furthermore, as our simple application shows, the impact of
this possible mistake may turn out to be dramatic and does not
vanish as the sample size increases.

Notice that similar issues have already been discussed in the
relevant literature
\cite{Lilliefors1967,Dyer1974,Stephens1974,GreenHegazy1976}. More
specifically, \cite{Stephens1976} shows that the mean of the
Anderson-Darling statistic shifts leftwards when the parameters of
the population distribution are unknown. Furthermore,
\cite{Stute1993,Rao2004} discuss the problem of approximating EDF
test statistics from a rather theoretical perspective. Yet, despite
the success of EDF-based GoF tests, no clear indications were given
-- to the best of our knowledge -- about the practical correct
Monte-Carlo procedure to be followed in order to approximate
test-statistic distributions in the case of unknown parameters. This
paper aims at shedding more light on the risks ensuing a wrong
specification of the Monte-Carlo simulation strategy, in all cases
where critical-value tables are not already available. Given the
lack of contributions addressing this topic, and the subtle nature
of the choice between Procedure A and B, our feeling is that
mistakes may be more likely than it may seem.

A final remark is in order. In our discussion we deliberately
focused only on the case where parameters to be estimated are
location and scale. In such an ``ideal'' situation, as we noted, the
distributions of the four EDF-based test-statistics that we have
considered do not depend on the true unknown parameters. Therefore,
in principle, to approximate their distributions one may generate,
in Step B1, a sufficiently large number of independently-drawn
N-sized samples from a $F(x;\theta^\ast)$, where $\theta^\ast$ is
any given value of the unknown parameters, and not necessarily their
empirical-sample estimates $\hat{\theta}(\underline{x}_N)$. Since
the distribution of the test is location- and scale-invariant, we
just need to make sure to apply Step B2 (i.e. re-estimation of
$\theta$ using $\underline{z}_N^j$) in order to avoid the implicit
assumption that parameters are known.

What happens if instead parameters are not location and scale but
are still unknown? In such a case, very common indeed (e.g., when
$F$ is a Beta or a Gamma distribution), test-statistic distributions
do depend on the true unknown parameter values
\cite{Darling1955,Durbin1975}. Therefore, Step B1 may be considered
as a first (good) guess towards the approximation of test-statistic
distributions. In fact, when parameters are not location and scale,
one cannot employ any given $\theta^\ast$ to generate Monte Carlo
samples. Since the ``true'' test-statistic distribution depends on
the ``true'' unknown parameter values, one would like to approximate
it with a sufficiently similar (although not exactly equal)
distribution, which can be easily obtained -- provided that
Procedure B is carried out -- by employing the empirical sample
estimates $\hat{\theta}(\underline{x}_N)$. In such a situation,
critical value tables are not typically available, because they
would depend on the empirical sample to be tested. Monte-Carlo
simulations are therefore required and choosing the correct
Procedure (B) instead of the wrong one (A) becomes even more crucial
than in the location - scale case.

\medskip
\begin{center}
\large \textbf{Acknowledgments}
\end{center}
\noindent This article enormously benefitted from comments and
suggestions made by Richard Lockhart, who patiently went through the
main points with his illuminating explanations. Thanks also to
Michael Stephens and Fabio Clementi for their very useful remarks
and suggestions. Any remaining errors are the sole responsibility of
the authors.

\newpage

\begin{table*} [h]\vspace*{12pt}
\centering
\begin{tabular}{cc|cc|cc|cc|cc}
\hline \hline
           &            & \multicolumn{ 2}{c|}{KS} & \multicolumn{ 2}{c|}{KUI} & \multicolumn{ 2}{c|}{CVM} & \multicolumn{ 2}{c}{AD2} \\

         N &          M &       Proc B &   Proc A &       Proc B &   Proc A &       Proc B &   Proc A &       Proc B &   Proc A \\
\hline \hline
&100&0.8220&1.2252&1.4235&1.6138&0.0690&0.2807&0.7194&2.2727\\
10&1000&0.8564&1.3035&1.4485&1.7270&0.0706&0.3707&0.6996&2.6139\\
&10000&0.8648&1.3482&1.4565&1.7314&0.0757&0.3800&0.7442&2.7667\\
\hline
&100&0.8159&1.4965&1.4376&1.7671&0.0985&0.5941&0.6863&3.1936\\
50&1000&0.8928&1.3813&1.4783&1.7715&0.1111&0.4512&0.7424&2.4106\\
&10000&0.8931&1.3623&1.4867&1.7489&0.1146&0.4446&0.7553&2.5414\\
\hline
&100&0.9380&1.4601&1.5831&1.7954&0.1308&0.4902&0.7664&2.7079\\
100&1000&0.8839&1.3525&1.5083&1.7162&0.1129&0.4634&0.7172&2.6332\\
&10000&0.8969&1.3587&1.4933&1.7407&0.1199&0.4478&0.7425&2.4740\\
\hline
&100&0.9177&1.2389&1.531&1.7857&0.1146&0.3455&0.7231&2.0889\\
500&1000&0.9125&1.3316&1.5139&1.7250&0.1273&0.4333&0.7769&2.3941\\
&10000&0.9108&1.3576&1.4998&1.7544&0.1261&0.4628&0.7543&2.5423\\
\hline
&100&0.9443&1.3723&1.4753&1.7879&0.1312&0.4116&0.7998&2.3795\\
1000&1000&0.9041&1.3858&1.5103&1.7814&0.1246&0.4674&0.7700&2.5817\\
&10000&0.9121&1.3606&1.5108&1.7575&0.1267&0.4578&0.7581&2.5076\\
\hline
&100&0.9689&1.2911&1.5656&1.8264&0.1405&0.4107&0.8099&2.2523\\
5000&1000&0.8944&1.3807&1.4801&1.7478&0.1204&0.4513&0.7180&2.4596\\
&10000&0.9116&1.3606&1.5120&1.7424&0.1274&0.4617&0.7613&2.5073\\\hline
\hline
\end{tabular}
\caption{Critical values at significance level $\alpha=0.05$ for the
four EDF tests considered. Proc A: Always using empirical-sample
estimates. Proc B: Parameters are re-estimated each time on
Monte-Carlo sample. KS=Kolmogorov-Smirnov; KUI=Kuiper;
CVM=Cram\'{e}r-Von Mises; AD2=Quadratic Anderson-Darling.}
\label{bigtable}
\end{table*}

\newpage
\clearpage

\end{document}